\documentclass[prb,twocolumn,showpacs,preprintnumbers,amsmath,amssymb]{revtex4}
\usepackage{graphicx}% Include figure files
\usepackage{dcolumn}% Align table columns on decimal point
\usepackage{bm}% bold math

\begin{document}

\newcommand*{\cm}{cm$^{-1}$\,}
\newcommand*{\Tc}{T$_c$\,}

%\reprint{APS/123-QED}

\title{Formation of partial energy gap below the structural phase transition
and the rare-earth element substitution effect on infrared phonons
in ReFeAsO (Re=La, Nd, and Sm)}% Force line breaks with \\
\author{T. Dong}
\author{Z. G. Chen}
\author{R. H. Yuan}
\author{B. F. Hu}
\author{B. Cheng}
\author{N. L. Wang}

\affiliation{Beijing National Laboratory for Condensed Matter
Physics, Institute of Physics, Chinese Academy of Sciences,
Beijing 100190, China}

%\date{March 26, 2008}% It is always \today, today,

\begin{abstract}

Single crystals of LaFeAsO, NdFeAsO, and SmFeAsO have been
prepared by means of a NaAs flux growth technique and studied by
optical spectroscopy measurements. We show that the spectral
features corresponding to the partial energy gaps in the
spin-density-wave (SDW) state are present below the structural
phase transition. This indicates that the electronic state below
the structural phase transition is already very close to that in
the SDW state. We also show that in-plane infrared phonon modes
display systematic shifts towards high frequency upon rare-earth
element substitutions for La, suggesting a strong enhancement of
the bonding strength. Furthermore, an asymmetric line-shape of the
in-plane phonon mode is observed, implying the presence of an
electron-phonon coupling effect in Fe-pnictides.

\end{abstract}



\maketitle

\section{Introduction}
The discovery of superconductivity at 26 K in F-doped
LaFeAsO\cite{Kamihara08} has created tremendous interests in the
scientific community. Shortly after this discovery, the
superconducting transition temperature T$_c$ was raised beyond 50
K through the substitution of La by rare-earth elements. T$_c$ is
found to be 41 K for F-doped CeFeAsO\cite{ChenPRLCe}, 52 K for
F-doped NdFeAsO\cite{Ren1} and 55 K for F-doped
SmFeAsO\cite{Ren2}. In the so-called 1111 structural type series,
the undoped parent compounds were commonly found to have a
spin-density-wave (SDW) ground state with a collinear
antiferromagnetic (AFM) spin configuration\cite{DongEPL,Clarina}.
A structural phase transition occurs prior to the magnetic
transition. Because superconductivity appears in the vicinity of
the magnetic ordered phase\cite{Clarina}, it is widely believed
that the spin fluctuations play a crucial role in the
superconducting pairing of electrons, while the electron-phonon
interactions could not explain the superconductivity in those
materials. A theoretical calculation even indicates that the
electron-phonon coupling could not lead to the superconductivity
with T$_c$ higher than 1 K in those materials\cite{Boeri}.

A critical question here is why the rare-earth element
substitutions for La in the 1111 series could significantly
enhance the superconducting transition temperature?  For La or
different rare-earth based parent compounds, the structural and
magnetic phase transitions were found to take place roughly at the
same temperatures, then the magnetic interactions should not have
much difference. On the other hand, the rare-earth element
substitutions lead to certain change in the lattice structure.
Some correlations between superconducting transition temperatures
and the Fe-As bond length or the Fe-As-Fe angle for the series
have been found\cite{JZhao,CHLee}. However, there is still a lack
of a complete understanding of the problem. In this work, we
performed optical spectroscopy study on single crystal samples of
the several different 1111 compounds, including LaFeAsO, NdFeAsO,
and SmFeAsO. We focus our attention on two issues. One is the
relationship between the structural and magnetic phase
transitions. We identify that the charge excitation gaps have
already opened at the structural phase transition, which then
discloses the intimate relation between the driving mechanisms of
the two transitions. The other is the effect of rare earth element
substitutions on the lattice modes. We found that the in-plane
phonon modes display systematic shifts towards high frequency with
the substitution of La by Nd, and Sm. We also observed asymmetric
line-shape of the in-plane phonon mode, providing direct evidence
for the presence of electron-phonon coupling. The results may
imply that the lattice vibrations play some role for the high
temperature superconductivity.

\section{Experiments and results}

\subsection{Experiments}

The crystal growth in 1111 systems has been proven to be
difficult. For a long period, only very small size single crystals
could be obtained with typical dimensions less than 300
$\mu$m.\cite{Zhigadlo,HSLee} Millimeter-sized single crystals were
obtained only by means of NaAs flux technique until
recently\cite{JQYan,ZGChen}. The ReFeAsO (Re=La, Nd, Sm) single
crystals used in this study were grown from such a technique, and
characterized by the X-ray diffractions (XRD) and dc resistivity
measurements. For the Sm-based sample, we put 10$\%$ F in the
initial composition. However the resultant crystals still have a
property of the parent compound. The major change is that the
resistivity anomaly corresponding to structural transition is
suppressed to about 120 K. Detailed descriptions about crystal
growth could be found in Ref. [13].

\begin{figure}[t]
\includegraphics[width=7cm]{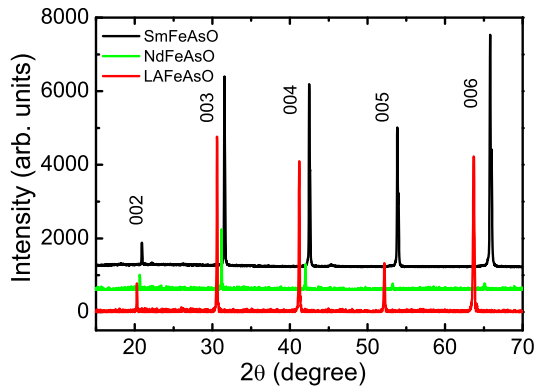}
\caption{(Color online) The \textit{(00l)} X-ray diffraction
patterns of single crystals ReFeAsO (Re=La, Nd, Sm).}
\end{figure}

Figure 1 shows the \textit{(00l)} XRD patterns for the single
crystal samples of ReFeAsO (Re=La, Nd, Sm) with Cu K$\alpha$
radiation. The XRD patterns indicate that the samples are of the
characteristic of good crystallization along the c axis. The
\textit{(00l)} peaks shift towards higher 2$\theta$ angles
systematically from LaFeAs to NdFeAsO and SmFeAsO, suggesting a
reduction of c-axis lattice parameter. The obtained c-axis lattice
parameter is c=8.758 $\AA$ for LaOFeAs, 8.595 $\AA$ for NdOFeAs,
and 8.497 $\AA$. The reduction of the c-axis lattice parameter is
apparently due to the smaller ionic radius of the rare earth
elements Nd and Sm. Those values are consistent with earlier
lattice parameters determined from the polycrystalline
samples.\cite{Pottgen}

Optical measurement was done on a Bruker IFS 66v/s spectrometer in
the frequency range from 40 to 25000 cm$^{-1}$. An \textit{in
situ} gold and aluminum overcoating technique was used to get the
reflectance \emph{R}($\omega$). The real part of conductivity
$\sigma_1(\omega)$ is obtained by the Kramers-Kronig
transformation of \emph{R}($\omega$).

\begin{figure}[t]
\includegraphics[width=8.5cm]{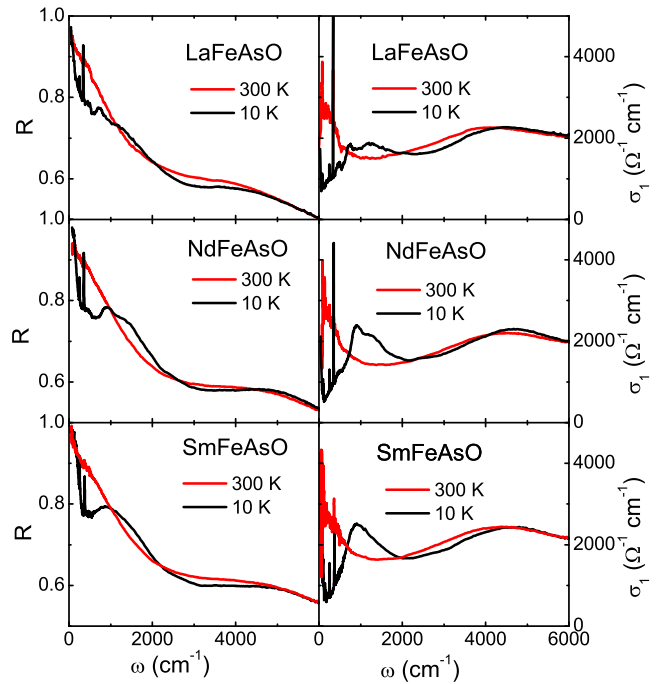}
\caption{(Color online) In-plane optical reflectance and
conductivity spectra up to 6000 \cm at 300 and 10 K for La-, Nd-
and Sm-based 1111 single crystal samples.}
\end{figure}

\subsection{The electronic spectra}

Figure 2 shows the optical reflectance and conductivity spectra up
to 6000 \cm of LaFeAsO, NdFeAsO and SmFeAsO at two representative
temperatures 300 K and 10 K, respectively. The data of LaFeAsO
were already presented in ref. 13 where its spectral features and
correlation effect were discussed. Here they are used for a
purpose of comparison. We can see that all the compounds exhibit
similar spectral features in both the paramagnetic and SDW states.
At room temperature, the reflectance R($\omega$) drops almost
linearly with frequency at low-$\omega$ region, then merges into
the high values of a background contributed mostly from the
interband transitions from the mid-infrared to visible regime. The
conductivity spectrum $\sigma_1(\omega$) displays a Drude-like
component at low frequency and a rather pronounced spectral weight
in the high frequencies. A broad peak is seen near 4500 \cm. In
the SDW state at 10 K, dramatic spectral change was seen at low
frequencies. The reflectance below 1000 \cm shows a remarkable
suppression, while it is enhanced between 1000 and 2000 \cm. Then,
in the conductivity spectra, the spectral weight is severely
suppressed at low frequencies and the missing spectral weight is
transferred to absorption peaks at high energies. This gives
optical evidence for the gap formation on the Fermi surfaces in
the SDW state. Since the reflectance still increases fast toward
unity at lower frequencies, residual free carriers or Drude
component is still left in the magnetic ordered state. Therefore,
the Fermi surfaces are not fully gapped below \emph{T}$_N$. Those
spectral features are also observed in 122-type AFe$_2$As$_2$
(A=Ba, Sr)\cite{Hu122} and 111-type Na$_{1-\delta}$FeAs single
crystals\cite{Hu111} across the SDW transitions, indicating the
generic properties of undoped FeAs-based systems.

As mentioned in the introduction, for 1111 parent compounds, a
structural transition from a tetragonal-to-orthorhombic phase
occurs prior to the SDW or magnetic transition, whereas the two
transitions occur simultaneously for 122-type compounds. The
relation between the structural and SDW transitions has been a
subject of many discussions\cite{Yildirim,Ma,Fang,Xu}.
Theoretically, it was suggested that the structural transition is
driven by the magnetic transition\cite{Yildirim,Ma,Fang,Xu}. Then
it is important to detect experimentally whether the charge
excitation gaps open in the structurally distorted phase (above
the magnetic transition) or not, an issue which has not been
addressed by any of previous optical measurements. For this
purpose, we present a detailed temperature-dependent R($\omega$)
and $\sigma_1(\omega$) spectra for one of the parent compounds,
NdFeAsO.

\begin{figure}[t]
\includegraphics[width=7.5cm]{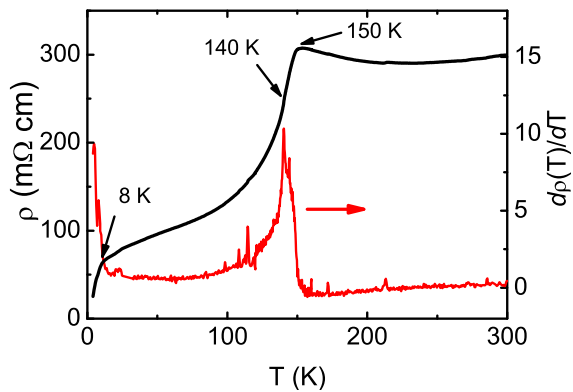}
\caption{(Color online) Temperature dependent resistivity
$\rho(T)$ and its derivative d$\rho(T)$/d\emph{T} curves for
NdFeAsO. The strong anomaly near 150 K is linked to the structural
transition, the sharp peak in d$\rho(T)$/d\emph{T} at 140 K
corresponds to the magnetic transition. Another decrease near 8 K
is attributed to the magnetic ordering of Nd moments.}
\end{figure}

Similar to other 1111-type compounds, NdFeAsO exhibits a
pronounced anomaly in the resistivity $\rho(T)$ near 150 K, below
which $\rho(T)$ decreases rapidly with decreasing
temperature.\cite{GFChen2} Figure 3 shows the temperature
dependent resistivity $\rho(T)$ curve of present single crystal
NdFeAsO. Compared with the data of polycrystalline
samples\cite{GFChen2}, the decrease of the resistivity near 150 K
is much faster for single crystal sample. Its temperature
derivative, d$\rho(T)$/d\emph{T}, shows a sharp peak at 140 K. It
is well known that the structural phase transition is responsible
for the dramatic change in the resistivity curve, while the
magnetic transition corresponds to the peak position in
d$\rho(T)$/d\emph{T}.\cite{McGuire} A neutron scattering
measurement on polycrystalline NdFeAsO sample indicates
T$_{SDW}$=141 K, being close to the peak temperature. The unusual
decrease at about 8 K could be attributed to the ordering of the
magnetic moment of Nd ions. The overall behavior of the
resistivity is very similar to the earlier report on
micrometer-sized NdFeAsO crystals\cite{PCheng}.

\begin{figure}
\includegraphics[width=7.5cm]{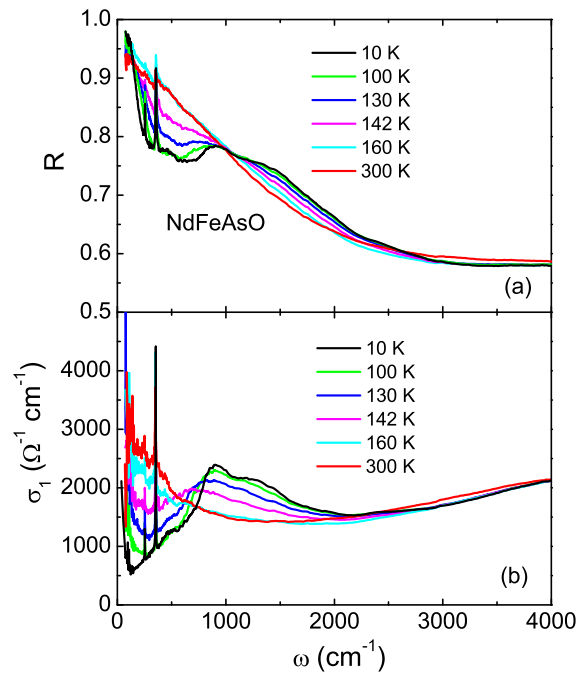}
\caption{(Color online) Reflectance and conductivity spectra of
NdFeAsO at different temperatures. As temperature decreases from
300 to 160 K, a narrowing of Drude-like component is seen.
However, a gap-like spectral weight suppression at low frequencies
is observed at 142 K, a temperature which is between the
structural and the magnetic transitions.}
\end{figure}

Figure 4 shows the temperature-dependent R($\omega$) and
$\sigma_1(\omega$) spectra of NdFeAsO. The energy gap features are
not present above the structural transition (see spectra above 160
K). However, below the structural phase transition, the low
frequency spectral weight suppression appears. The spectral
features observed at 142 K, a temperature between the structural
and magnetic transitions, resemble to that at very low
temperatures in the SDW state, although the features are weak at
such a high temperature. This clearly indicates that the charge
excitation gaps start to open at the structural phase transition,
thus demonstrating that the electronic state below but near the
structural phase transition is already very close to that in the
magnetic state. The observation would imply essentially the same
driving mechanism for both transitions. Although the long-range
magnetic order is not formed at the structural transition,
short-range order, either nematic order or very strong
spin-fluctuations, would exist between the structural and magnetic
transition. We note that, a recent ARPES experiment on a 111-based
system NaFeAs, which also has separated structural and magnetic
transitions, revealed band-folding effect just below the
structural transition\cite{He}; while a NMR study on NaFeAs
indicated strong enhancement of magnetic fluctuations\cite{Yu}.
Both are consistent with our optical measurement on 1111 systems.

\subsection{The phonon spectra}

Despite the fact that the electronic spectra display similar
features for La or different rare-earth substituted 1111 parent
compounds, we found that there exists a systematic frequency shift
of infrared active in-plane phonon modes, which is the focus in
the rest of the present paper. Figure 5 shows the low frequency
conductivity spectra of the ReFeAsO (Re=La, Nd, Sm) crystals at 10
K, where two phonon modes could be seen in the frequency range. We
shall show later that those phonon peaks do not show clear
frequency shift across the structural and magnetic transitions.

\begin{figure}
\includegraphics[width=8cm]{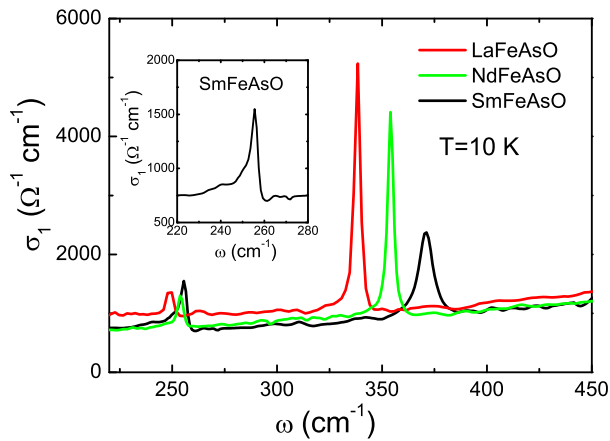}
\caption{(Color online) In-plane infrared phonon modes at 10 K for
La-, Nd-, and Sm-based 1111 single crystals. Both the Fe-As and
O-derived modes show hardening with the rare-earth element
substitutions. Inset shows the expanded frequency region for Fe-As
mode of SmFeAsO crystal. An asymmetric behavior is clearly seen,
indicating strong electron-phonon coupling.}
\end{figure}

It is known that ReFeAsO crystallizes in the tetragonal
\emph{P}\textit{4/nmm} structure at high temperature. According to
symmetry analysis, there are six infrared active modes
3\emph{A}$_{2u}$ +3\emph{E}$_u$ along the \emph{c}-axis and in
\emph{ab} planes, respectively.\cite{Hadjiev} Early infrared
reflectance measurement on polycrystalline LaFeAsO sample
indicates five infrared active modes in the far-infrared
region\cite{DongEPL}, which could be assigned to those
\emph{A}$_{2u}$ and \emph{E}$_u$ modes. Here we only observe two
\emph{E}$_u$ modes at 248 and 339 \cm for LaFeAsO in the \emph{ab}
plane response. It is noted that the 248 \cm mode seen in
polycrystalline samples was assigned to the \emph{A}$_{2u}$ mode
along the c-axis in early studies, while another mode with
slightly higher frequency at 266 \cm was assigned to the in-plane
\emph{E}$_u$ mode\cite{Yildirim,Marini}. A comparison of the
phonon modes between the single crystal and the polycrystalline
samples indicates that the assignments of the two modes should be
reversed. The 248 \cm \emph{E}$_u$ mode involves the in-plane
displacement of Fe-As ions. The 339 \cm mode, which is
particularly strong in intensity, is associated with the in-plane
displacement of the O-La(Re) bonds. Its high energy location
indicates that this mode is mostly oxygen
derived\cite{Singh237003}. Several interesting features about the
phonon modes are found from Fig. 5: (1) both in-plane \emph{E}$_u$
modes shift to higher frequencies in Nd- and Sm-based samples. The
shift is more pronounced for the oxygen-derived mode; (2) The
strength of Fe-As mode increases from LaFeAsO to NdFeAsO and
SmFeAsO, while the intensity of the O-Re (Re=La, Nd, Sm) mode
decreases; (3) There exist asymmetric line-shapes for the phonon
modes (Fano line shape\cite{Fano}), evidencing sizable
electron-phonon coupling effect. An expanded frequency plot for
the Fe-As mode of SmFeAsO sample is shown in the inset of Fig. 5.

The increase of the phonon frequency in rare-earth element
substituted systems strongly indicates an enhancement of the
in-plane Fe-As and O-Re bonding strengths. With the substitution
of La by Nd and Sm, the masses of the rare-earth ions increase. If
there was no bonding strength change, we would expect a reduction
of frequencies of the in-plane O-Re stretching mode. The
noticeable increase of the mode frequency suggests a reduction of
the bond-length, and it overcomes the effect of the mass change of
the rare-earth ions. Not only the O-derived mode shows a
substantial increase, but the in-plane Fe-As mode also shows a
hardening from La to Nd and Sm substitutions, although less
pronounced. It is worth noting that the Raman scattering
measurements also indicate a strong hardening of the Raman active
O mode\cite{Hadjiev,Marini}. Overall, the observations are
consistent with the structural characterization data which show a
decreasing of the Fe-As bond length\cite{JZhao,CHLee}.The above
result is not affected by the F-doping, since low level F-doping
is found to have little influence on the phonon frequencies of the
parent compound\cite{L.Zhang}.

\begin{figure}
\includegraphics[width=8cm]{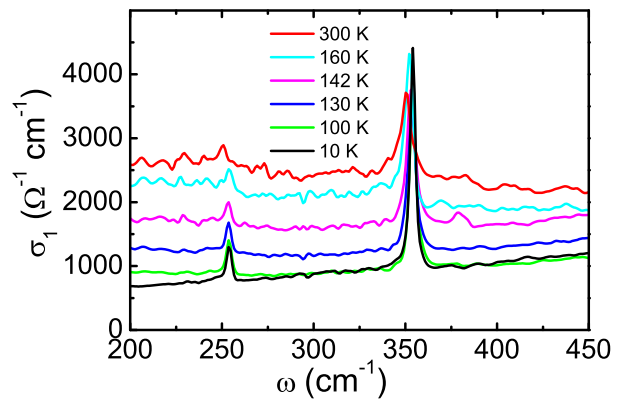}
\caption{(Color online) The temperature dependence of the in-plane
phonon modes of NdFeAsO single crystal.}
\end{figure}

Figure 6 shows the temperature evolution of the \emph{E}$_u$
phonon modes across the structural and magnetic transitions for
the prototype NdFeAsO crystal. Similar to the case of infrared
measurement on the polycrystalline samples,\cite{DongEPL} we do
not observe any splitting or new phonon lines below the structural
transition. According to density function theory calculations on
LaFeAsO, the splitting of the infrared active phonon modes would
be very small and therefore could not be well resolved by the
optical measurement.\cite{Yildirim} The intensities of the two
modes show significant temperature dependence. They are strongly
enhanced below the structural or magnetic transition. This
behavior is similar to the observation in BaFe$_2$As$_2$ crystal
where two infrared active in-plane modes near 94 and 253 \cm are
observed, and the 253 \cm mode shows a strong intensity
enhancement below the structural transition. A natural explanation
is that the screening effect from the conduction electrons is
significantly reduced due to the partial gapping of Fermi
surfaces. Alternatively, it was proposed that the intensity change
could be due to a charge redistributions at each atom leading to a
change in bonding between different atoms.\cite{Akrap} This point
of view might be useful in interpreting the intensity increase of
the Fe-As stretching mode but the decrease of the O-La mode from
LaFeAsO to NdFeAsO and SmFeAsO.

The above experimental results indicate a good correlation between
the bonding length in the crystal structure and the frequencies of
the infrared phonon modes. Since the rare-earth element
substitutions for La significantly enhance the superconducting
transition temperature in doped compounds, it is of worth to
further investigate the role played by the electron-phonon
coupling for the superconductivity.

\section{Conclusions}

We have grown single crystals of the parent compounds LaFeAsO,
NdFeAsO, and SmFeAsO by means of a NaAs flux growth technique and
studied their charge excitations and infrared phonon spectra by
optical spectroscopy measurement. We identify that the charge
excitation gaps are already present below the structural phase
transition with spectral feature similar to those found at the SDW
state, thus illustrating that the electronic state below the
structural phase transition is essentially the same as that in the
SDW state. We show that in-plane infrared phonon modes display
noticeable shifts towards high frequency upon rare-earth element
(Nd, Sm) substitutions for La, indicating a strong enhancement of
the bonding strength. The study also yields evidence for the
presence of electron-phonon coupling effect in the compounds.

\begin{acknowledgments}
We thank P. Dai, J. L. Luo and T. Xiang for useful discussions.
This work was supported by the National Science Foundation of
China, the Knowledge Innovation Project of the Chinese Academy of
Sciences, and the 973 project of the Ministry of Science and
Technology of China.

\end{acknowledgments}

%\bibliography{MgIrB}% Produces the bibliography vi BibTeX.

\end{document}